\documentclass[conference]{IEEEtran}
\usepackage{cite}
\usepackage{amsthm,amsmath,amssymb,amsfonts}
\usepackage{algorithmic}
\usepackage{graphicx}
\usepackage{textcomp}
\usepackage{orcidlink}
\usepackage{threeparttable}
\usepackage{xcolor}
\usepackage{booktabs}
\usepackage[caption=false]{subfig}
\newtheorem{safety}{Safety Property}
\newtheorem{liveness}{Liveness Property}
\def\BibTeX{{\rm B\kern-.05em{\sc i\kern-.025em b}\kern-.08em
    T\kern-.1667em\lower.7ex\hbox{E}\kern-.125emX}}
\begin{document}

\title{TrustMesh: A Blockchain-Enabled Trusted Distributed Computing Framework for Open Heterogeneous IoT Environments
}

\author{
\IEEEauthorblockN{Murtaza Rangwala\,\orcidlink{0009-0003-4578-8671} and Rajkumar Buyya\,\orcidlink{0000-0001-9754-6496}}
\IEEEauthorblockA{\textit{Cloud Computing and Distributed Systems (CLOUDS) Laboratory} \\
\textit{School of Computing and Information Systems} \\
\textit{The University of Melbourne, Australia} \\
Email: mrangwala@student.unimelb.edu.au, rbuyya@unimelb.edu.au}
}

\maketitle

\begin{abstract}
The rapid evolution of Internet of Things (IoT) environments has created an urgent need for secure and trustworthy distributed computing systems, particularly when dealing with heterogeneous devices and applications where centralized trust cannot be assumed. This paper proposes TrustMesh, a novel blockchain-enabled framework that addresses these challenges through a unique three-layer architecture combining permissioned blockchain technology with a novel multi-phase Practical Byzantine Fault Tolerance (PBFT) consensus protocol. The key innovation lies in TrustMesh's ability to support non-deterministic scheduling algorithms while maintaining Byzantine fault tolerance, features traditionally considered mutually exclusive in blockchain systems. The framework supports a sophisticated resource management approach that enables flexible scheduling decisions while preserving the security guarantees of blockchain-based verification. Our experimental evaluation using a real-world cold chain monitoring scenario demonstrates that TrustMesh successfully maintains Byzantine fault tolerance with fault detection latencies under 150 milliseconds, while maintaining consistent framework overhead across varying computational workloads even with network scaling. These results establish TrustMesh's effectiveness in balancing security, performance, and flexibility requirements in trustless IoT environments, advancing the state-of-the-art in secure distributed computing frameworks.
\end{abstract}


\section{Introduction}
\label{sec:introduction}

Rapid advances in the Internet of Things (IoT) and distributed computing paradigms have brought to the forefront critical challenges in creating secure, trustable, and auditable systems. As the complexity and scale of IoT ecosystems and their critical applications increase, ensuring data integrity, maintaining trust, and providing stringent auditability have become paramount concerns, particularly in environments where centralized trust cannot be assumed \cite{lockl_toward_2020}.

This becomes particularly evident in specific application domains where trustless operations and auditability are critical. In smart city traffic management systems, data from multiple sources must be processed and shared without the risk of manipulation by any single entity. Healthcare IoT, including wearable devices and remote patient monitoring systems, requires a trustless framework to maintain patient privacy while allowing authorized access to critical health data. Similarly, in industrial IoT settings, predictive maintenance and supply chain management across multi-vendor environments demand tamper-proof data integrity and transparent audit trails to ensure operational reliability and regulatory compliance.

These challenges are compounded by the inherent heterogeneity of IoT environments, which manifests in multiple dimensions, including substantial variations in the capabilities of distributed nodes. These disparities span computational power, architectural design of processors, memory constraints, and diversity in supported communication protocols \cite{merlino_enabling_2019}. Furthermore, the wide array of IoT applications introduce additional complexity, with varying operational units, inter-dependencies, and resource demands. Some applications consist of independent operations, while others require sequential execution or involve complex inter-dependencies. Additionally, these applications impose varying demands on the system, with some requiring intensive computational resources and others prioritizing rapid response times.

To address these multifaceted challenges, this paper proposes TrustMesh, a blockchain-enabled distributed computing framework for trustless and heterogeneous IoT environments. Our proposed framework introduces a novel decentralized architecture that combines a permissioned blockchain with a multi-phase Practical Byzantine Fault Tolerance (PBFT) consensus based protocol for scheduling decisions. This unique approach allows for the use of non-deterministic scheduling algorithms, including genetic algorithms, machine learning, and heuristics, while maintaining the benefits of blockchain technology. The main contributions of this paper include: 
\begin{itemize}
    \item A systematic analysis of existing distributed computing frameworks for IoT, with emphasis on their approaches to trust, resource heterogeneity, and operational auditability.
    \item The design and implementation of TrustMesh\,\footnote{Available at: https://doi.org/10.5281/zenodo.14677989}, a decentralized framework addressing these challenges.
    \item A novel secure protocol for resource management and task scheduling that enables flexible resource allocation while maintaining Byzantine fault tolerance.
    \item An experimental evaluation of the framework's resource utilization, performance scalability and reliability using a 21-node heterogeneous testbed in a cold chain monitoring scenario.
\end{itemize}

The rest of the paper is organized as follows: Section~\ref{sec:relatedwork} provides a detailed background and related work in distributed computing for IoT and decentralized systems. Section~\ref{sec:architecture} presents the architecture and design principles of TrustMesh, including an in-depth explanation of the multi-phase commit protocol used for scheduling decisions. Section~\ref{sec:formalproof} presents a formal verification and correctness analysis of the proposed protocol. Section~\ref{sec:implementation} describes the implementation details. Section~\ref{sec:performanceeval} presents our experimental setup and evaluation results. Section~\ref{sec:discussion} discusses the implications of our findings. Section~\ref{sec:threats} addresses the threats to validity and limitations of our approach. Section~\ref{sec:conclusion} concludes the paper along with directions for future work.

\section{Related Work}
\label{sec:relatedwork}

\begin{table*}[t]
\centering
\caption{Qualitative Comparison of Related Works}
\label{tab:comparison}
\begin{tabular}{|l|c|c|c|c|c|c|c|c|}
\hline
\textbf{Work} & \textbf{\begin{tabular}[c]{@{}c@{}}Complete\\Decentralized\\Control\end{tabular}} & \textbf{\begin{tabular}[c]{@{}c@{}}Heterogeneous\\Multi-Application\\Workflows Support\end{tabular}} & \textbf{\begin{tabular}[c]{@{}c@{}}App Lifecycle\\Management\end{tabular}} & \textbf{\begin{tabular}[c]{@{}c@{}}Fault\\Tolerance\end{tabular}} & \textbf{\begin{tabular}[c]{@{}c@{}}Dynamic\\Resource\\Discovery\end{tabular}} & \textbf{\begin{tabular}[c]{@{}c@{}}Authentication, \\Encryption \&\\Integrity\end{tabular}} & \textbf{\begin{tabular}[c]{@{}c@{}}Practical\\Implementation\end{tabular}} \\
\hline
\cite{tuli_fogbus_2019} & $\times$ & $\times$ & $\times$ & $\circ$ & \checkmark & $\circ$ & \checkmark \\
\hline
\cite{yuan_coopedge_2021} & \checkmark & $\times$ & $\times$ & \checkmark & $\times$ & \checkmark & \checkmark \\
\hline
\cite{lei_groupchain_2020} & \checkmark & $\times$ & $\times$ & \checkmark & $\times$ & \checkmark & \checkmark \\
\hline
\cite{kumar_blockedge_2020} & $\times$ & $\times$ & $\times$ & $\circ$ & \checkmark & \checkmark & $\times$ \\
\hline
\cite{mayer_fogchain_2021} & \checkmark & $\times$ & $\times$ & \checkmark & $\times$ & $\circ$ & $\times$ \\
\hline
\cite{nunez-gomez_hidra_2021} & \checkmark & $\times$ & \checkmark & \checkmark & \checkmark & $\circ$ & \checkmark \\
\hline
TrustMesh & \checkmark & \checkmark & \checkmark & \checkmark & \checkmark & \checkmark & \checkmark \\
\hline
\end{tabular}
\label{relatedworks}
\begin{tablenotes}
      \small
      \item \checkmark: Full support \hspace{0.5cm} $\circ$: Partial support \hspace{0.5cm} $\times$: No support
\end{tablenotes}
\end{table*}

The evolution of distributed computing frameworks for IoT applications has been fundamentally shaped by the emergence of fog computing and subsequent developments in blockchain technologies and distributed systems design. Bonomi et al. \cite{bonomi_fog_2014} introduced the foundational concepts of fog computing, establishing it as a hierarchical distributed platform that extends cloud capabilities to the network edge. Since then, several works have explored the critical requirements for latency reduction, bandwidth optimization, and efficient resource allocation in fog environments \cite{yi_survey_2015}.

While these advancements addressed performance and resource management challenges, the distributed nature of fog computing introduced significant challenges in establishing trust and ensuring security across heterogeneous nodes. The lack of centralized control raised concerns about data integrity, authentication, and secure resource sharing among untrusted parties. The integration of blockchain with fog computing systems emerged as a promising solution to these challenges. 

Tuli et al. \cite{tuli_fogbus_2019} proposed FogBus, implementing one of the first blockchain-based frameworks for fog computing. FogBus introduced concepts for secure data logging and transaction verification, but its reliance on a partially centralized control structure limited its applicability in truly distributed environments. The framework architecture struggled with scalability issues in large-scale deployments. This limitation was addressed by Yuan et al. \cite{yuan_coopedge_2021} through CoopEdge, which introduced a reputation-based consensus mechanism and a permissioned blockchain network. CoopEdge's novel approach to trust establishment and maintenance significantly improved system reliability, though its focus on computation offloading left gaps in comprehensive application lifecycle management.

Lei et al. \cite{lei_groupchain_2020} proposed Groupchain, introducing a two-chain structure and leader group consensus mechanism to improve transaction throughput. While this approach enhanced transaction processing, it lacked comprehensive features for system administration and operational oversight. Kumar et al. \cite{kumar_blockedge_2020} proposed BlockEdge, implementing a hybrid blockchain architecture that combines permissioned and permissionless chains across its three-tier structure. BlockEdge's approach to chain management demonstrated strong potential for heterogeneous environments, though its implementation remained limited to simulations, leaving questions about real-world performance unanswered.

More recent frameworks evolved to provide more comprehensive solutions. Mayer et al. \cite{mayer_fogchain_2021} presented FogChain, specifically targeting healthcare data management through fog computing. Their work demonstrated significant latency improvements through its fog-based architecture, particularly in handling sensitive medical data with strict privacy requirements. Núñez-Gómez et al. \cite{nunez-gomez_hidra_2021} presented HIDRA, demonstrating practical implementation of a blockchain-based architecture on resource-constrained devices. HIDRA's implementation on actual hardware provided valuable insights into the challenges of deploying blockchain solutions on edge devices, though with limited scalability testing.

TrustMesh builds upon these foundations while addressing their limitations through a novel combination of permissioned blockchain technology, comprehensive task scheduling, and efficient resource management. By introducing an innovative consensus-based protocol for scheduling decisions, TrustMesh overcomes the deterministic constraints that limit existing blockchain-based solutions. Table~\ref{relatedworks} identifies and compares some key features of related frameworks with TrustMesh. Unlike previously proposed works that focused on specific aspects of fog and edge computing, TrustMesh provides a comprehensive solution that addresses security, performance and management concerns in a unified architecture.

\section{TrustMesh: Architecture and Design Principles}
\label{sec:architecture}

\begin{figure}[htbp]
\centering
\includegraphics[width=1\columnwidth]{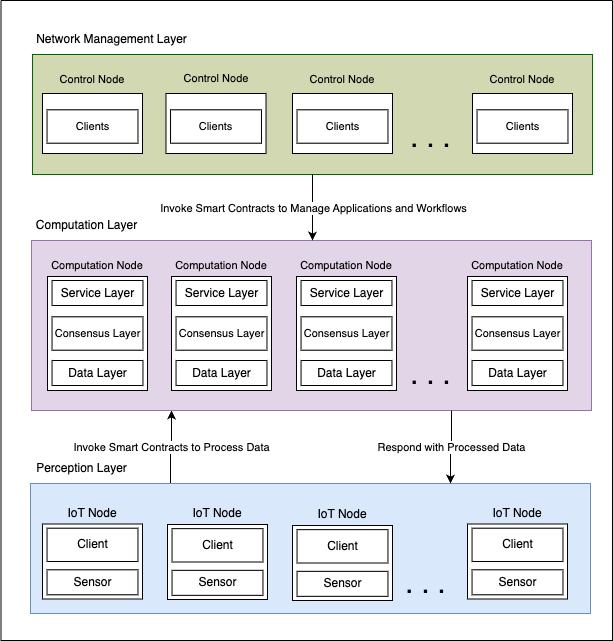} 
\caption{TrustMesh High Level Architecture}
\label{highlevelarchitecture}
\end{figure}

TrustMesh employs a multi-tiered architecture comprising three primary layers that work in concert to provide secure, decentralized IoT data processing. Fig.~\ref{highlevelarchitecture} presents a high-level architectural overview of the framework, illustrating the relationships between these layers and their key components. In the following subsections, we examine each layer's architecture and interaction mechanics in detail.

\subsection{Network Management Layer}
\label{subsec:nmlayer}

The Network Management Layer forms the administrative backbone of the TrustMesh framework, providing essential infrastructure for system setup, configuration, and application management. This layer consists of control nodes that, while not participating directly in data processing, serve as critical access points for network administrators to manage applications and define workflows within the network. Control nodes in this layer host two primary clients for application deployment and workflow management.

The Application Deployment Client facilitates the deployment and management of applications across the network. It provides administrators with the capability to deploy new application images, instantiate containers from existing images, restart failed application instances, and remove applications or their images from the network. When an administrator initiates an action, the client checks the locally hosted registry within the Network Management Layer and pushes the application image if not already present. Subsequently, a smart contract is invoked in the Computation Layer. This smart contract stores the action details and when applicable, the digital signature of the application image, in the blockchain. Once stored, the smart contract emits an event that is broadcast to all the compute nodes in the network. Each compute node is equipped with an event handler that listens for these broadcasts. Upon receiving an event, if the action type is application deployment, the event handler verifies the image signature stored in the blockchain and then pulls the application image from the registry to local storage. For container instantiation requests, the handler creates new containers from the previously verified local image using the specified configuration parameters. For container termination, it gracefully stops and removes the targeted containers. In the case of deployment removal, the handler terminates all associated containers and removes the application image from local storage. This event-driven architecture ensures consistent and secure application management across the entire network, maintaining the integrity of the distributed system.

The Workflow Management Client allows for the creation of complex workflow pipelines. Administrators can define dependency graphs representing the relationships between multiple applications deployed on the network. These workflows are represented as directed acyclic graphs (DAGs), where each node corresponds to a pre-existing application on the network, and edges represent the flow of data or execution order. This representation enables the definition of complex, interdependent processing pipelines with varying levels of parallelism and sequentiality. Once created and stored in the blockchain, these workflows become immutable and guide the processing of data received from the Perception Layer. The flexibility of this workflow structure allows TrustMesh to accommodate a wide range of application scenarios, from simple linear processes to complex, branching computations.

A key architectural decision in TrustMesh is the separation between the Network Management Layer and the underlying network infrastructure. While the network is deployed using a Kubernetes cluster, application and workflow management processes operate independently from the Kubernetes control plane. This architectural decision enables distributed control of applications and workflows through redundant control nodes, significantly enhancing system reliability. The segregation also creates distinct security boundaries, containing potential breaches within the application management domain rather than exposing the entire infrastructure. Although control nodes introduce some degree of centralization to the system, they serve solely as interfaces for triggering smart contracts and are only required for initial setup and management operations. This design ensures that the ongoing operation of deployed applications and workflows remains unaffected by control node availability, thereby avoiding critical points of failure. To ensure operational integrity, the architecture implements blockchain-based smart contracts that govern all management actions. These contracts, deployed across compute nodes, facilitate secure communication between the Management and Computation layers through robust encryption protocols. Furthermore, each management operation is cryptographically signed and recorded on the blockchain, establishing an immutable audit trail that proves particularly valuable in multi-stakeholder environments as it ensures accountability without requiring trust in any single party's record-keeping.

\subsection{Computation Layer}
\label{subsec:computationlayer}

The Computation Layer serves as the foundational element of TrustMesh, comprising compute nodes that may be edge, fog, or cloud entities participating in the blockchain network. This layer is organized into three key components: the data sub-layer, consensus sub-layer, and service sub-layer, each designed to address specific requirements of trustless, distributed environments.

\subsubsection{Data Sub-Layer}
\label{subsubsec:datasublayer}
This sub-layer manages persistent and temporary storage requirements through three primary storage mechanisms. The first is a distributed database cluster implementing a multi-master architecture with peer-to-peer data replication. This architecture enables efficient write operations as each node maintains a local copy of persistent data, with replication managed independently by the database cluster. This persistent storage primarily handles IoT data from the perception layer when data persistence is explicitly enabled for the transaction, providing granular control over data retention.

The second component is the blockchain's immutable ledger, maintained locally by each compute node. This ledger records data movement across the network and ensures data integrity through digital signatures. By storing these signatures in the immutable ledger, the system enables verification of data stored in the persistent storage cluster while maintaining a transparent audit trail of all operations.

The third component is a distributed cache cluster implemented as an in-memory sharded system based on hash slot partitioning. Unlike the persistent storage, this cache cluster prioritizes access speed over complete data replication across nodes. While configured with redundancy measures, the cache cluster primarily facilitates inter-node communication within the computation layer, managing temporary data storage for ongoing operations.

\subsubsection{Consensus Sub-Layer and Protocol Design}
\label{subsubsec:consensussublayer}

\begin{figure*}[htbp]
\centering
\includegraphics[width=1\textwidth]{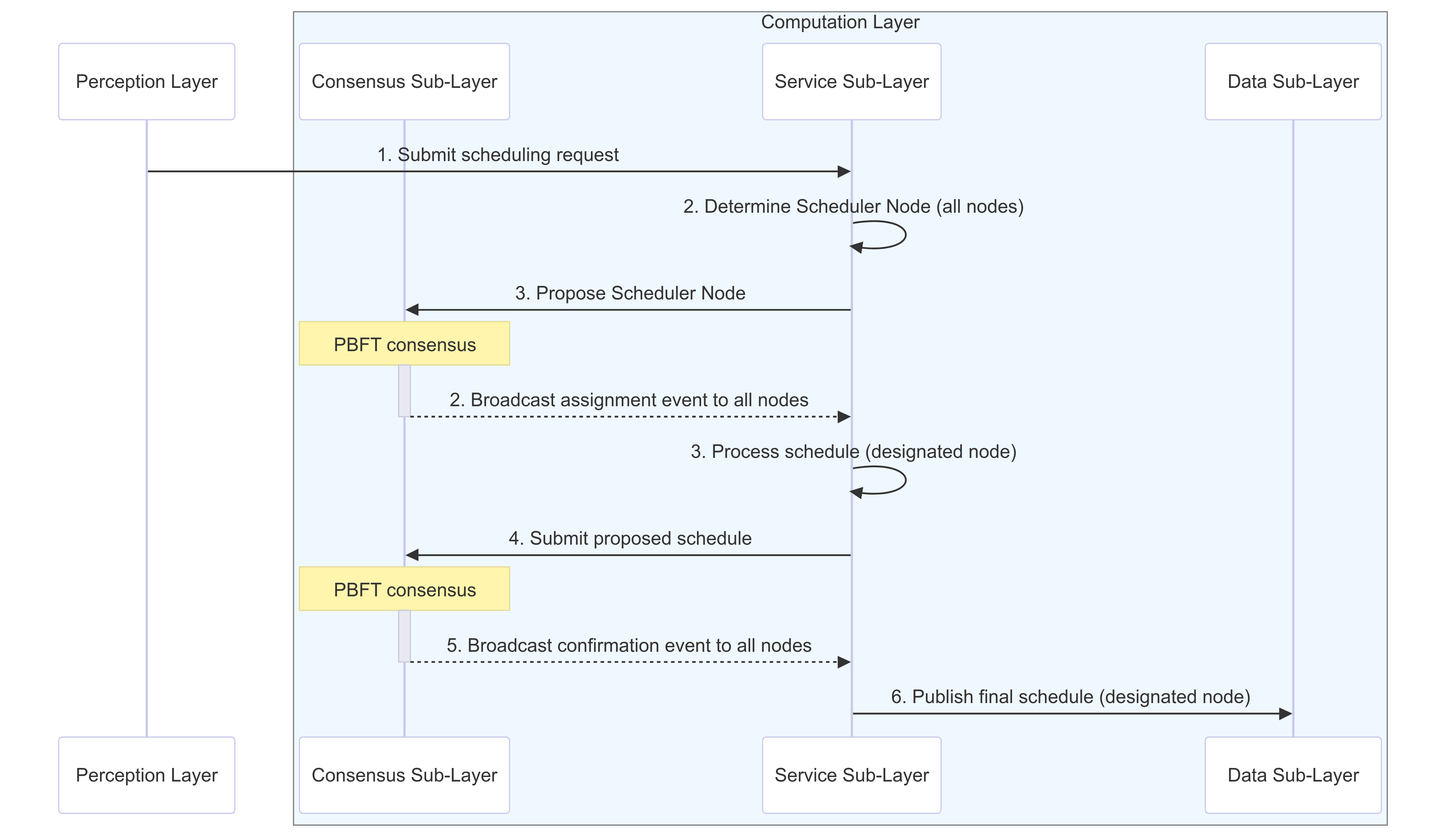} 
\caption{Multi-Phase Commit Protocol}
\label{multiphasecommitprocessflow}
\end{figure*}

This sub-layer employs PBFT for all operations except scheduling. The original PBFT protocol, proposed by Castro, Liskov et al. \cite{castro_practical_1999}, enables distributed systems to maintain decentralized control even when up to one-third of the network nodes exhibit Byzantine behavior. While this approach effectively ensures consistency across standard workflow operations, it proves unsuitable for scheduling decisions that rely on non-deterministic methods, such as those driven by machine learning algorithms or other meta-heuristics.

To overcome this challenge, TrustMesh extends PBFT with a novel protocol, as illustrated in Fig~\ref{multiphasecommitprocessflow}. To formally define this protocol, we first establish the system model. Let $\mathcal{S}$ represent the set of all possible system states. Each state $s \in \mathcal{S}$ comprises:
\begin{itemize}
\item A set of nodes $N = \{n_1, ..., n_k\}$
\item A schedule request $r$ with identifier $id_r$
\item A schedule proposal $p$ with $status_p \in \{Requested, Active, Finalized, Failed\}$
\item A blockchain state $B$
\item Resource state $R$ mapping each node to its available resources
\end{itemize}

The protocol consists of three distinct phases:
\begin{enumerate}
\item {
Request and Designation Phase ($\phi_1$):
\begin{equation}
\phi_1: (s_0, r, R) \rightarrow (s_1, n_d)
\end{equation}
where $s_1$ updates $B$ with the following transactions:
\begin{itemize}
\item Data transaction: Contains the IoT data signature.
\item Status transaction: Represents the initial schedule state as $Requested$.
\item Schedule request transaction: Includes the designated node information.
\end{itemize}
and the designated node $n_d$ is determined by the deterministic function $f$, where:
\begin{equation}
f(N, R) = n^* \text{ where } n^* = \max_{n \in N} (w_c \cdot c_n + w_m \cdot m_n)
\end{equation}
with $c_n$ and $m_n$ representing available CPU and memory resources for node $n$, and $w_c$, $w_m$ being weighting factors.
}
    \item {
    Generation Phase ($\phi_2$):
\begin{equation}
\phi_2: (s_1, n_d) \rightarrow (s_2, p)
\label{eq:statephase2}
\end{equation}
where $n_d$ generates schedule $p$ using a customizable scheduling algorithm after receiving the schedule-request event.
    }
    \item {
    Confirmation Phase ($\phi_3$):
\begin{equation}
\phi_3: (s_2, p) \rightarrow s_3
\label{eq:statephase3}
\end{equation}
where $s_3$ updates $B$ with the following transactions:
\begin{itemize}
\item Status transaction: Represents the schedule state as now $Active$.
\item Schedule Confirmation Transaction: Includes the schedule $p$ finalized by $n_d$.
\end{itemize}
and PBFT consensus is achieved on the proposed schedule through a validation process incorporating both protocol-level and network-defined criteria. Participating nodes evaluate the schedule against a set of customizable validation rules $V = \{v_1, ..., v_m\}$, which can include constraints on vendor distribution, resource utilization, geographic placement, and security clearance levels. A schedule $p$ is considered valid if and only if:
\begin{equation}
\forall v_i \in V : v_i(p) = true
\label{eq:validation}
\end{equation}
The schedule is added to the ledger and published to the cache cluster only after satisfying all validation criteria and achieving PBFT consensus, enabling networks to implement context-specific security policies while maintaining Byzantine fault tolerance.
    }
\end{enumerate}

The generation phase as defined in \eqref{eq:statephase2} permits the designated node to implement any scheduling algorithm $\mathcal{A}: \mathcal{S} \rightarrow \mathcal{P}$ that maps from the system state space $\mathcal{S}$ to the space of permissible schedules $\mathcal{P}$. In our implementation, we use a Least-Connected Dynamic Weighted Round Robin (LCDWRR) approach. To formally describe this algorithm, we first define its operational model.

Let $\mathcal{G} = (V, E)$ be the workflow dependency graph where:
\begin{itemize}
\item $V$ is the set of tasks
\item $E$ represents dependencies between tasks
\item $level: V \rightarrow \mathbb{N}$ maps each task to its topological level
\item $N$ is the set of available nodes
\item $R_n$ represents resource state of node $n \in N$
\end{itemize}

The LCDWRR algorithm must produce a schedule $p \in \mathcal{P}$ satisfying these properties:
\begin{enumerate}
    \item {
    Resource Feasibility:
\begin{equation}
\begin{split}
\forall t \in V, \exists n \in N: & \; cpu_{req}(t) \leq cpu_{avail}(n) \; \land \\
& \; mem_{req}(t) \leq mem_{avail}(n)
\end{split}
\end{equation}
    }
    \item {
    Level Preservation:
\begin{equation}
\forall (t_i, t_j) \in E: level(t_i) < level(t_j)
\end{equation}
    }
    \item {
    Schedule Completeness:
\begin{equation}
\forall t \in V: \exists! n \in N: t \in schedule(n)
\end{equation}
    }
\end{enumerate}

Our LCDWRR implementation achieves these properties through:
\begin{enumerate}
    \item {
    Topological sorting:
\begin{equation}
levels = \{l: \{v \in V | level(v) = l\}\}
\end{equation}
    }
    \item {
    Node eligibility based on resources:
\begin{equation}
\begin{split}
N_e(t) = \{n \in N \;|\; & \; cpu_{req}(t) \leq cpu_{avail}(n) \; \land \\
& \; mem_{req}(t) \leq mem_{avail}(n)\}
\end{split}
\end{equation}
    }
    \item {
    Probabilistic node selection:
\begin{itemize}
\item Node load: 
\begin{equation}
\begin{split}
load(n) = & \; 0.5 \cdot \frac{cpu_{used}}{cpu_{total}} \; + \\
& \; 0.5 \cdot \frac{mem_{used}}{mem_{total}}
\end{split}
\end{equation}
\item Selection weight: 
\begin{equation}
w_n = \frac{1}{load(n) + 0.1}
\end{equation}
\item Selection probability: 
\begin{equation}
\begin{split}
P(n|t) &= \frac{w_n}{\sum_{i \in N_e} w_i} 
\end{split}
\end{equation}
\end{itemize}
    }
\end{enumerate}

\subsubsection{Service Sub-Layer}
\label{subsubsec:servicesublayer}
This sub-layer drives the operational logic of compute nodes through a combination of event handlers, scheduled processes, and subscribers that collectively facilitate workflow execution. Central to this functionality are two key event handlers: the Application Image Event Handler and the Schedule Event Handler. The Application Image Event Handler, as explained in Section~\ref{subsec:nmlayer}, processes application management events initiated by network management actors. Meanwhile, the Schedule Event Handler oversees schedule generation within the computation layer, leveraging a pluggable Schedule class. This class encapsulates customizable scheduling algorithms, enabling the framework to address varying requirements across diverse IoT environments.

It also includes a resource registration process that executes at configurable intervals to monitor compute node resources. This process maintains resource data at three levels: latest state in the cache cluster for immediate access, complete historical records in the persistent data storage, and cryptographic signatures in the blockchain for verification of data integrity. To optimize blockchain performance and reduce network load, these signatures are not recorded individually. Instead, they are aggregated into fixed-size batches and committed to the blockchain at intervals determined by the resource update frequency and batch size pre-configured by the network administrator.

Finally, the task executor within this layer forms the cornerstone of workflow management, orchestrating data processing according to predefined workflow specifications. When a scheduling request is initiated from a source IoT node, it is first recorded in the blockchain. Upon schedule generation and publication to the cache cluster's pub/sub channel, all participating nodes are notified of their assigned tasks and their respective topological execution levels. The Task Executor implements a sophisticated dependency resolution mechanism that ensures tasks are executed only when their prerequisites are satisfied.

Tasks at level 0, having no dependencies, begin execution immediately upon schedule notification. Each node executes its assigned tasks and stores the output in the cache cluster, with optional persistence to the persistent storage cluster. Upon task completion, nodes publish completion notifications to the pub/sub channel, enabling the coordinated progression of the workflow. Tasks at the same level can execute in parallel once their respective dependencies are satisfied, enhancing throughput without compromising workflow integrity. Nodes independently verify completion status of prerequisite tasks, fetch required input data from the cache cluster, and proceed with execution. This parallel execution capability enhances system throughput while maintaining data consistency through the pub/sub notification system.

For workflow completion, the Task Executor implements a finalization mechanism. The last node to complete its task at the highest level verifies the completion status of all workflow tasks. Upon confirmation, it updates the workflow status in the blockchain to $Finalized$ and securely transmits the final output to the source IoT node using ZeroMQ. This communication is encrypted using the source node's public key, shared during the initial scheduling request.

\subsection{Perception Layer}
\label{subsec:perceptionlayer}

The Perception Layer serves as the interface between the TrustMesh framework and the physical environment, with IoT nodes functioning as the primary actors within this layer. These nodes fulfill a dual role, operating both as producers of raw data and consumers of processed results.

This layer's architecture implements a zero-process baseline approach, whereby IoT nodes maintain no running processes by default upon framework deployment. This minimalist design is fundamental to accommodating resource-constrained IoT devices. The framework's functionality in this layer is delivered through two primary helper components, while the specific implementations for raw data management and result consumption are left to the framework users, allowing processes to be added based on particular use cases, resource constraints, and environmental requirements.

The first helper component, the Transaction Initiator, performs the essential function of data preparation. It accepts raw data as input and systematically encapsulates it into transaction batches suitable for transmission to the Computation Layer for processing. The second component, the Response Manager, implements cryptographic operations using Curve25519 for secure communication. It generates public-private key pairs and establishes a network socket for receiving processed data. The public key is transmitted to the Computation Layer within the transaction payload, enabling secure response delivery. The Response Manager then utilizes the corresponding private key to decrypt incoming data at the established socket, making it available for consumption by the IoT node.

These components are implemented as libraries, providing flexibility for integration into use-case specific client implementations. This architectural decision enables adaptability while maintaining the framework's core security and data handling capabilities across diverse IoT applications.

\section{Multi-Phase Commit Protocol: Verification and Correctness}
\label{sec:formalproof}

The multi-phase commit protocol requires formal verification to ensure correctness and fault tolerance. We analyze the protocol's safety and liveness properties.

\subsection{Safety Properties}
\label{subsec:safetyproperties}

\begin{safety}[Deterministic Designation]
For a given resource state $R$ and set of nodes $N$, the designation function $f$ always produces the same designated node $n_d$.
\end{safety}

\begin{IEEEproof}
The designation function $f$ implements a deterministic algorithm that sorts nodes by resources, uses atomic operations for consistent views, applies fixed weighting, and breaks ties consistently. Therefore, given input state $(N, R)$, $f$ will always produce identical output $n_d$.
\end{IEEEproof}

\begin{safety}[Agreement]
For any schedule request $r$, if a node accepts schedule $p$ as Active, no other node can accept a different schedule $p'$ for $r$.
\end{safety}

\begin{IEEEproof}
Assume by contradiction that schedules $p$ and $p'$ become $Active$ for request $r$. This would require two blocks with confirmation transactions achieving PBFT consensus. This is impossible because:
\begin{enumerate}
\item The smart contract uses $id_r$ in state addressing
\item PBFT ensures total order of dependent transactions
\item State validation prevents multiple confirmations for the same $id_r$
\end{enumerate}
\end{IEEEproof}

\begin{safety}[Validity]
An Active schedule $p$ must originate from the designated node $n_d$.
\end{safety}

\begin{IEEEproof}
This property is enforced through multiple verification layers:
\begin{enumerate}
\item Designation: Deterministic selection and blockchain recording with PBFT consensus
\item Proposal: Identity verification and restricted processing to designated node
\item Confirmation: Smart contract validation of proposer identity against blockchain record
\item Consensus: Network-wide validation of designation rules
\end{enumerate}
\end{IEEEproof}

\subsection{Liveness Properties}
\label{subsec:livenessproperties}

\begin{liveness}[Termination]
Under synchrony, a valid schedule request $r$ eventually results in an Active schedule $p$.
\end{liveness}

\begin{IEEEproof}
Given synchrony assumptions, termination is guaranteed through eventual resource consistency, bounded designation time, reliable event delivery via blockchain subscription, PBFT confirmation, and reliable schedule distribution.
\end{IEEEproof}

\subsection{Fault Tolerance Analysis}
\label{subsec:faulttoleranceanalysis}
The protocol inherits Byzantine fault tolerance from PBFT, tolerating up to $f$ Byzantine nodes where $n \geq 3f + 1$ ($n$ represents total node count). Correctness is maintained through deterministic designation, PBFT validation, timeout-based reassignment, and fault-tolerant schedule distribution.

\section{Implementation}
\label{sec:implementation}

This section presents a comprehensive analysis of TrustMesh's implementation specifics, emphasizing the system's containerization strategy, blockchain integration, and data management approach. 

\subsection{Container Orchestration and Network Deployment}
\label{subsec:container-orchestration-impl}
The physical implementation of the TrustMesh network employs K3S, a lightweight Kubernetes distribution optimized for IoT and edge computing environments. The selection of K3S is influenced by its comprehensive support for both ARM64 and ARMv7 architectures, addressing the inherent heterogeneity of IoT and edge computing hardware. Furthermore, K3S leverages a single binary packaging approach to achieve exceptional resource efficiency among lightweight Kubernetes distributions. Comparative analysis with microK8S, K0S, and Microshift demonstrates K3S's lower idle state overhead while maintaining high control plane responsiveness~\cite{koziolek_lightweight_2023}.

This low resource overhead allows the network to leverage Kubernetes' inherent capabilities for high availability through automated failover, system resilience through self-healing mechanisms, and enhanced security through role-based access control. The deployment workflow separates core framework services from data management components, with independent deployments for persistent database and cache clusters enabling granular resource allocation and scaling.

\subsection{Blockchain Implementation and Smart Contracts}
\label{subsec:blockchain-impl}
TrustMesh implements its blockchain functionality using Hyperledger Sawtooth, a permissioned blockchain platform that provides the security and performance characteristics required for trustless IoT environments. As demonstrated in \cite{wang_performance_2020}, Sawtooth achieves superior performance metrics with throughput of 500-2000 TPS and latency of 0.5-5 seconds, compared to Ethereum's 10-30 TPS with 5-second latency and Fabric's 100-200 TPS with 1-10 second latency in comparable environments. The platform's modular design facilitates independent smart contract development and pluggable consensus mechanisms, enabling adaptation to varying IoT environment requirements. The framework's smart contracts are implemented using the Sawtooth Python SDK, chosen for Python's capabilities in rapid prototyping and testing.

\subsection{Data Storage Architecture}
\label{subsec:data-store-impl}
The data management infrastructure implements a hybrid approach utilizing CouchDB for persistent storage and Redis for in-memory operations. CouchDB's multi-master architecture provides robust support for complex data aggregation through Map-Reduce views, while its conflict resolution mechanisms effectively handle concurrent updates in distributed environments~\cite{anderson_couchdb_2010}. The document-oriented nature of CouchDB particularly suits the semi-structured nature of IoT data, enabling efficient storage and retrieval patterns.

Redis complements this through its sharded implementation, facilitating high-performance temporary storage operations and pub/sub messaging for event distribution~\cite{carlson_redis_2013}. The combination of these technologies enables atomic operations crucial for maintaining consistency in distributed operations while supporting the eventual consistency model appropriate for IoT environments.

\subsection{Integration and Communication}
\label{subsec:communication-impl}
System integration is achieved through a sophisticated combination of REST APIs, ZeroMQ messaging, Redis pub/sub channels, and native client libraries. This comprehensive integration strategy ensures efficient inter-component communication while maintaining system decentralization and security requirements. The implementation leverages ZeroMQ's inherent support for secure messaging patterns~\cite{hintjens_using_nodate} for point-to-point communication, Redis pub/sub channels for efficient data distribution across computation nodes, while REST APIs provide standardized interfaces for data operations and service communication.

All communication channels implement robust security measures: SSL/TLS encryption for REST interfaces and data store interactions, and Curve25519 cryptography for ZeroMQ messaging services, ensuring secure data exchange across the distributed system.

\section{Performance Evaluation}
\label{sec:performanceeval}

\subsection{Experimental Setup}
\label{subsec:experimentalsetup}

\begin{table*}[t]
\renewcommand{\arraystretch}{1.3}
\caption{Experimental Testbed Configuration}
\label{table:testbed}
\centering
\begin{tabular}{p{2.55cm} p{1.5cm} p{0.6cm} p{3.5cm} p{5cm}}
\toprule
\textbf{Layer} & \textbf{Node Type} & \textbf{Qty} & \textbf{Hardware Configuration} & \textbf{Operating Systems} \\
\midrule
Network Management & Control Node & 1 & 8 CPU cores, 32GB RAM, 30GB storage & Single Distribution$^*$ \\
\midrule
Computation & Compute Node & 16 & Single core, 4GB RAM, 30GB storage & AlmaLinux 8 (2), Debian 10/11/12 (6), Fedora 38/39/40 (4), Ubuntu 20.04/24.04 (4) \\
\midrule
Perception & IoT Node & 4 & Single core, 4GB RAM & Debian 12 (2), Ubuntu 20.04 (2) \\
\bottomrule
\multicolumn{5}{l}{\footnotesize $^*$Note: Production environments should implement multiple control nodes for enhanced reliability}\\
\end{tabular}
\end{table*}

The experimental evaluation of TrustMesh utilized a heterogeneous testbed of 21 nodes, provisioned from the Melbourne Research Cloud, across three framework layers as detailed in Table~\ref{table:testbed}. The Computation Layer comprised 16 single-core nodes (4GB RAM, 30GB storage) running diverse Linux distributions including AlmaLinux, Debian, Fedora and Ubuntu to reflect the heterogeneous nature of edge computing environments. Network latency measurements were conducted between all possible pairs of compute nodes, with 10 measurements taken for each pair. The resulting mean latency of 5.12 milliseconds aligned with typical values observed in edge computing environments \cite{zhang_how_2023}. The average latency between IoT nodes in the Perception Layer and the Computation Layer was measured at 12.2 milliseconds. The Perception Layer employed four single-core IoT nodes (4GB RAM) running Debian and Ubuntu, while the Network Management Layer and docker registry operated on an 8-core control node (32GB RAM, 30GB storage), though production environments would typically implement redundant control nodes.

\begin{figure}[htbp]
\centering
\includegraphics[width=1\columnwidth]{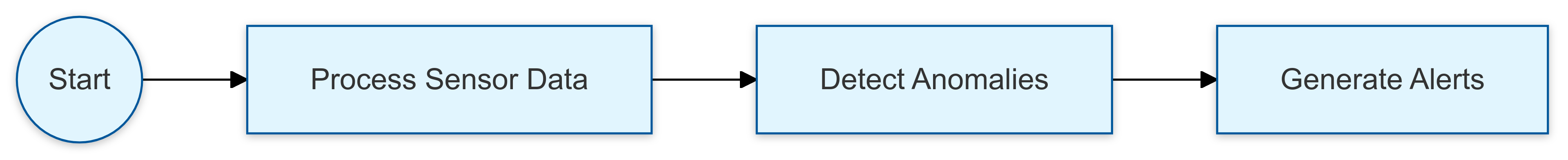} 
\caption{Cold-Chain Monitoring DAG}
\label{cold-chain-dag}
\end{figure}

The performance evaluation employed a cold chain monitoring scenario with a three-stage workflow (sensor processing, anomaly detection, and alert generation) arranged sequentially as shown in Fig~\ref{cold-chain-dag}, demonstrating TrustMesh's practical application in industrial IoT environments requiring continuous temperature and moisture monitoring.

\subsection{System Resource Utilization}
\label{subsec:systemresourceutil}

\begin{figure*}[!t]
    \centering
    \subfloat[CPU Usage by Workflow Progression]{%
        \includegraphics[width=0.48\textwidth]{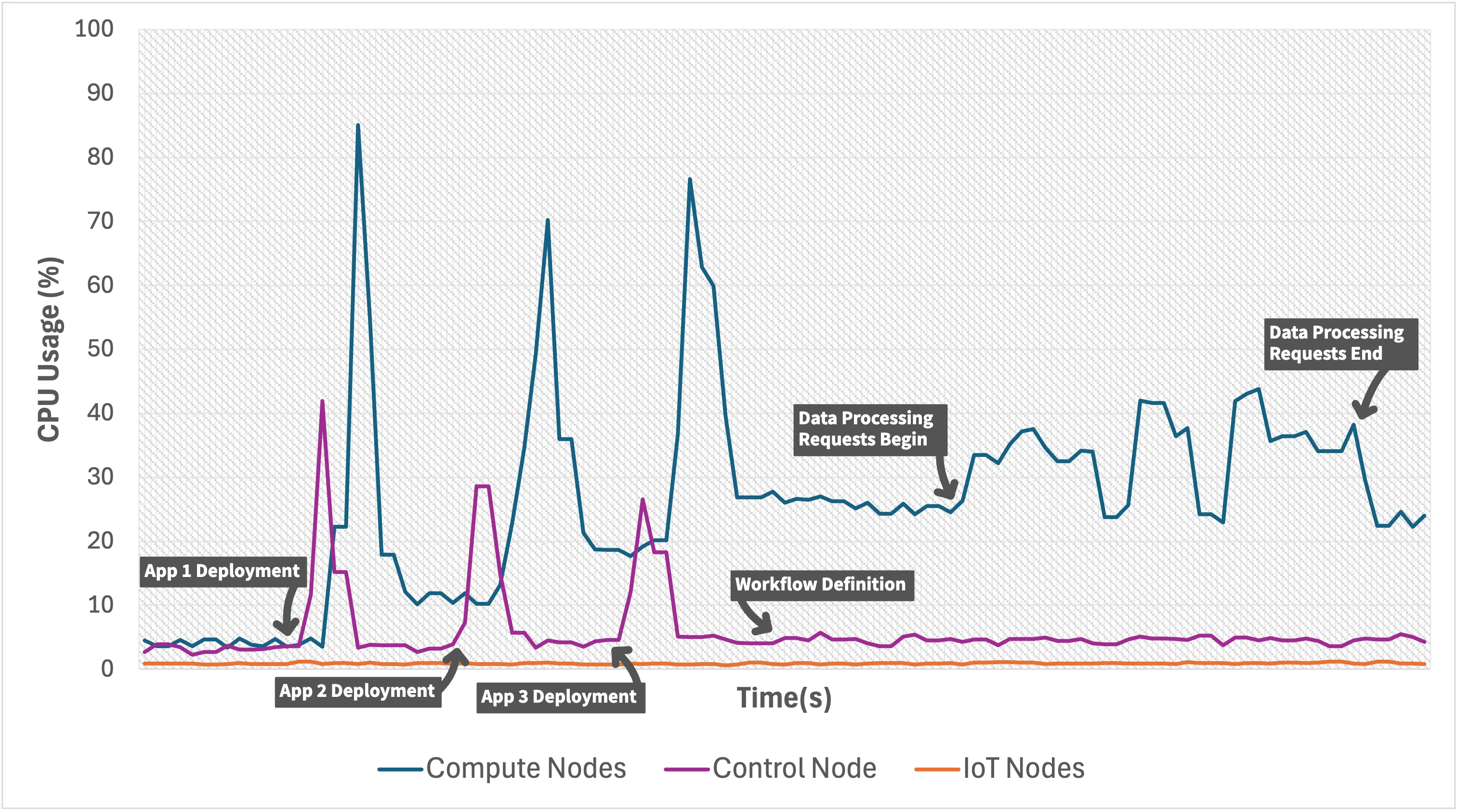}
        \label{fig:cpu-usage}
    }
    \hfil 
    \subfloat[RAM Usage by Workflow Progression]{%
        \includegraphics[width=0.48\textwidth]{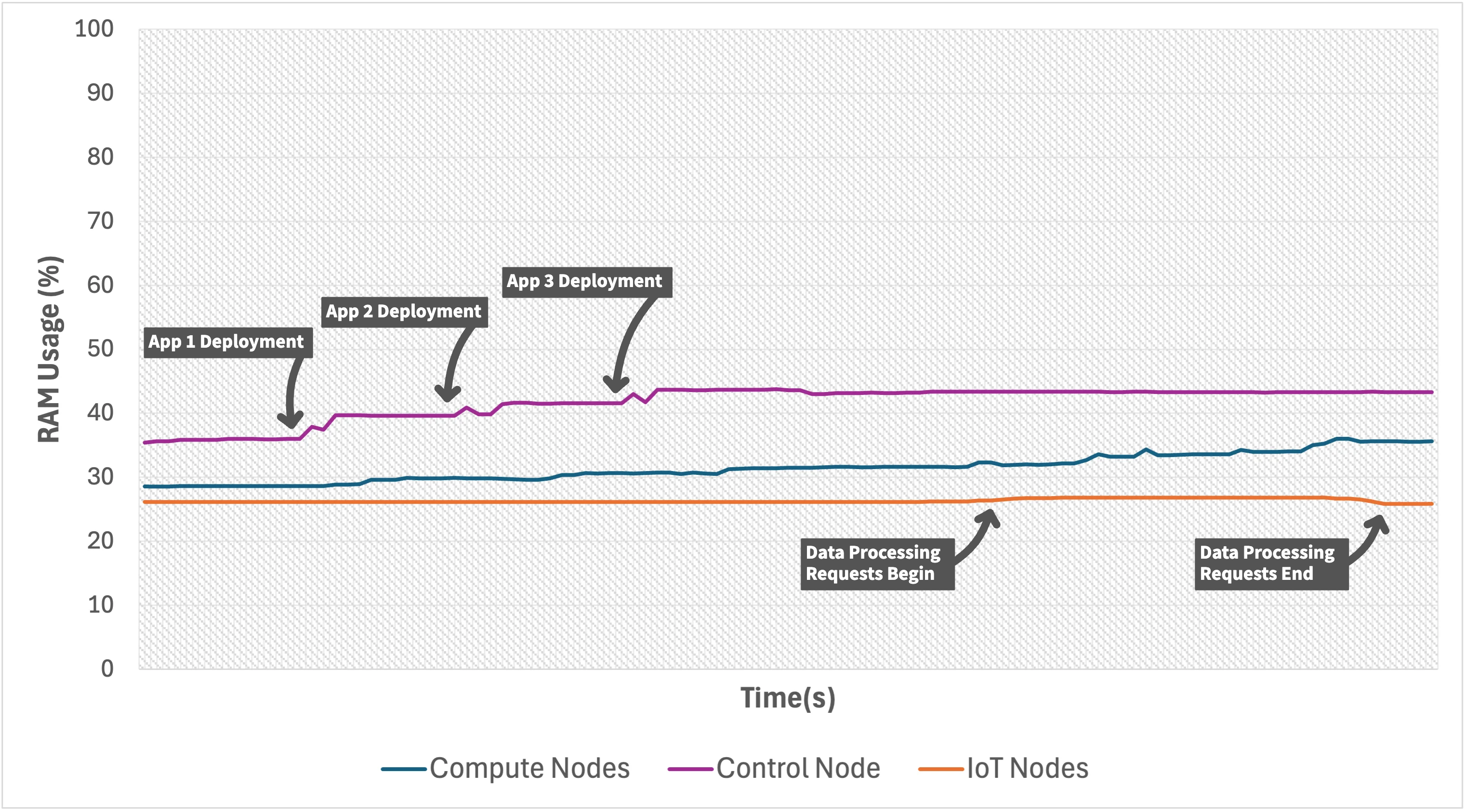}
        \label{fig:ram-usage}
    }
    \caption{Resource Usage}
    \label{fig:resource-usage}
\end{figure*}

We conducted comprehensive resource utilization analysis across all three architectural layers during the execution of the cold-chain monitoring workflow. The CPU and RAM utilization patterns, illustrated in Fig~\ref{fig:resource-usage}, demonstrate distinct characteristics for each layer throughout different stages of execution.

The CPU utilization profile exhibits three prominent peaks corresponding to the deployment of the workflow's constituent applications. These peaks occur in a staggered pattern - first appearing in the Network Management Layer, followed by corresponding peaks in the Computation Layer approximately 30 seconds later. Post-application deployment, the workflow definition stage shows minimal computational overhead in the Network Management Layer, with no significant CPU spikes observed. The system transitions into its operational phase during the latter portion of the experiment, marked by a sustained elevation in Computation Layer CPU utilization to approximately 45\%. This elevation coincides with the initiation of concurrent data processing requests from the four IoT nodes, operating at 30-second intervals. The Perception Layer maintains consistently low CPU utilization throughout the experiment.

Memory utilization patterns reveal a distinct stepping pattern in the Network Management Layer, with three clear incremental increases in RAM usage corresponding to each application deployment phase. The Computation Layer exhibits a similar but less pronounced stepping pattern during deployment, followed by a notable increase in RAM consumption as it enters the operational phase. In contrast, the Perception Layer maintains stable baseline consumption throughout, with minimal variance between idle and operational states. A slight increase in IoT node RAM usage is observed during active request transmission.

To ensure statistical validity, measurements represent averages across all nodes within each layer, sampled at 10-second intervals, with the complete workflow executed across ten independent iterations to mitigate potential sampling biases.

\subsection{Scalability and Performance}
\label{subsec:scalabilityandperformance}

\begin{figure}[htbp]
\centering
\includegraphics[width=1\columnwidth]{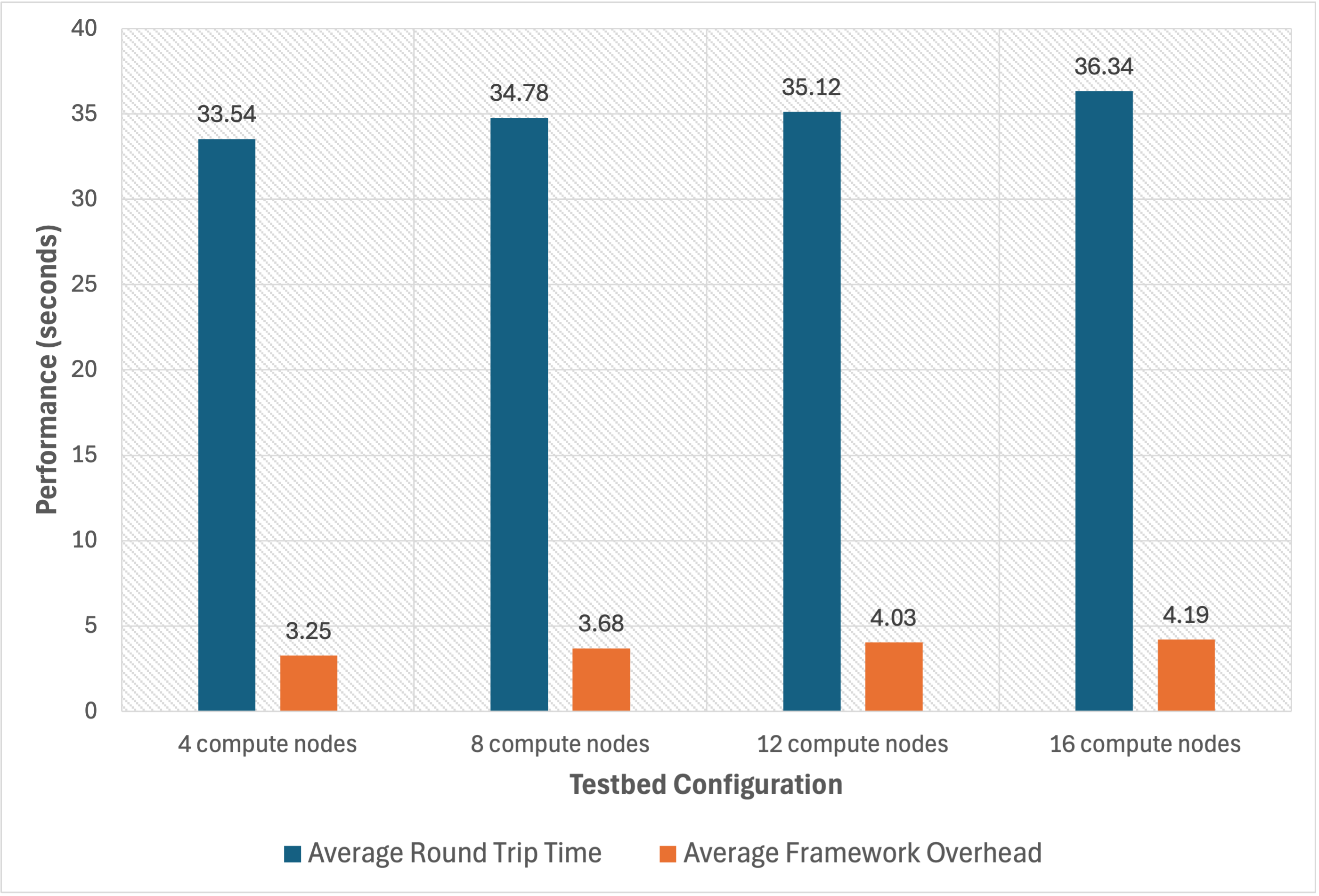} 
\caption{Performance Comparison by Testbed Configuration}
\label{performancebytestbedconfig}
\end{figure}

To evaluate the framework's scalability characteristics, we conducted extensive performance testing across four distinct testbed configurations, varying the number of compute nodes from 4 to 16 while maintaining a single control node and four IoT nodes throughout all tests. The evaluation focused on two critical metrics: Request Round Trip (RRT) time, measuring the complete cycle from initial request to result delivery, and Framework Overhead (FO) time, calculated as the differential between RRT and actual application execution duration. For statistical validity, each configuration was tested with 30 independent data processing requests at 30 second intervals from each node in the perception layer simultaneously.

The empirical results, illustrated in Fig~\ref{performancebytestbedconfig}, show the RRT times increasing from 33.54 seconds with 4 compute nodes to 36.34 seconds with 16 nodes, while the FO increased from 3.25 seconds to 4.19 seconds across the same scaling range.

We conducted additional testing of the FO time across three variations of the anomaly detection application within the cold-chain scenario with varying computational requirements (light, medium, and intensive processing). The results revealed consistency in FO times within each of the four network configurations. For instance, in the 16-node configuration, the FO time remained at approximately 4.2 seconds across all three workload variations.

\subsection{Byzantine Fault Tolerance Analysis}
\label{subsec:BFT}

To evaluate the framework's resilience against Byzantine behavior, we conducted controlled experiments simulating two attack scenarios within the 16-node computation layer, with 5 nodes (31.25\% of the network) configured to accept malicious scheduling attempts. The first scenario tested unauthorized schedule generation, where a non-designated node attempted to propose schedules violating the third safety property in Section~\ref{sec:formalproof}. The second scenario examined schedule request interference, where nodes designated for a new request attempted to propose schedules for requests with existing valid schedules.

\begin{table}[ht]
\caption{Byzantine Fault Tolerance Performance Metrics}
\label{tab:bft_metrics}
\centering
\begin{tabular}{lcc}
\hline
\textbf{Metric} & \textbf{Scenario 1} & \textbf{Scenario 2} \\
\hline
Detection Latency (ms) & 127 ± 15 & 142 ± 18 \\
Recovery Time (s) & 5.23 ± 0.12 & 5.31 ± 0.15 \\
CPU Utilization (\%) & 45.5 & 47.8 \\
\hline
\end{tabular}
\end{table}

The system detected and rejected unauthorized attempts during the confirmation phase with 100\% accuracy and zero false positives, triggering new schedule requests after the pre-configured 5-second failure timeout period. Table~\ref{tab:bft_metrics} presents key performance metrics averaged over 100 test iterations. Detection latency represents the time between malicious proposal and network detection, while recovery time measures the duration until a new valid schedule is created. The computation layer's CPU utilization during attack scenarios measured 45.5\% and 47.8\% for scenarios 1 and 2 respectively.

\section{Discussion}
\label{sec:discussion}

The performance evaluation results reveal several important characteristics of TrustMesh's behavior in heterogeneous edge environments. The staggered deployment peaks directly reflect the framework's two-phase deployment mechanism: applications must first be pushed to the control node's local registry before compute nodes can pull and instantiate them. While this introduces a 30-second deployment delay, it ensures reliable application distribution and prevents resource contention during deployment.

The framework's performance scaling characteristics warrant particular attention. The modest 8.3\% increase in RRT despite quadrupling compute nodes suggests efficient consensus operation at scale. More significantly, the consistency in Framework Overhead across varying computational workloads reveals a key architectural strength: the overhead is predominantly determined by consensus operations rather than application complexity. As a result, the measured overhead remains relatively constant within each configuration as long as the total workload requirements remain within the network's processing capacity.

The Byzantine fault tolerance results demonstrate a critical advancement in securing decentralized edge computing. The framework's ability to maintain correctness with up to one-third Byzantine nodes while supporting non-deterministic scheduling represents a significant capability. The detection latencies under 150 milliseconds indicate minimal impact on normal operations, while the consistent recovery times around 5 seconds suggest reliable system stabilization following attack detection.

These findings have important implications for edge computing deployments. The framework's ability to maintain consistent overhead across varying workloads while supporting Byzantine fault tolerance makes it particularly suitable for environments where security requirements must be balanced against performance constraints. TrustMesh's architecture inherently trades some performance overhead for enhanced security and auditability through its blockchain-based consensus mechanisms, making it well-suited for IoT applications requiring strict regulatory compliance, such as healthcare monitoring or supply chain management. The cold chain monitoring results validate this design choice, demonstrating that TrustMesh can effectively support complex industrial IoT workflows in environments where security and auditability take precedence over raw performance metrics.

\section{Threats to Validity}
\label{sec:threats}

\subsection{Internal Validity}
\label{subsec:internal-validity-threats}
The primary threat to internal validity stems from the measurement methodology employed in our performance evaluation. Resource utilization measurements could be influenced by background processes despite our controlled environment. To mitigate this threat, we conducted multiple iterations of each experiment and reported averaged results. Additionally, we utilized dedicated monitoring tools to isolate the resource consumption of TrustMesh components from system processes.

The selection of the 30-second interval for data processing requests in scalability testing could affect the observed performance patterns. This interval was chosen to balance data collection frequency with battery optimization, as IoT devices in typical supply chain use-cases are battery-operated. Moreover, the scalability testing involved simultaneous data processing requests from four IoT nodes, better reflecting real-world deployment scenarios. While different intervals might yield varying performance characteristics, we partially addressed this through our additional testing of different computational workloads, which demonstrated consistent FO across varying processing intensities.

\subsection{External Validity}
\label{subsec:external-validity-threats}
Our experimental testbed, while heterogeneous, represents a specific subset of possible edge computing environments. The use of 21 nodes with predefined hardware configurations may not capture all deployment scenarios in production environments. However, we carefully selected diverse Linux distributions and hardware profiles to represent common edge computing deployments. The cold chain monitoring scenario, while representative of industrial IoT applications, may not encompass all possible use cases for blockchain-enabled distributed computing.

The Byzantine fault tolerance evaluation considered two specific attack scenarios. While these scenarios represent common attack patterns, they may not cover all possible Byzantine behaviors in production deployments. To mitigate this limitation, we selected attack scenarios that target core consensus mechanisms, as these represent fundamental security challenges in distributed systems.

\subsection{Construct Validity}
\label{subsec:construct-validity-threats}
The chosen metrics (Request Round Trip time and Framework Overhead) may not capture all aspects of framework performance. To address this threat, we complemented these primary metrics with detailed resource utilization measurements and Byzantine fault tolerance metrics. The separation of Framework Overhead from application execution time provides a more precise measure of the framework's impact, though this separation may not be as distinct in all deployment scenarios.

The relationship between node count and performance metrics observed in our experiments appears to show linear scaling behavior within the tested range (4-16 nodes). However, this observation should be interpreted with caution, as PBFT consensus theoretically exhibits $O(n^2)$ time complexity \cite{castro_practical_1999}. Larger deployments would likely demonstrate this quadratic scaling characteristic more prominently.

\section{Conclusion and Future Work}
\label{sec:conclusion}

This paper proposed TrustMesh, a blockchain-enabled distributed computing framework designed for trustless heterogeneous IoT environments. The framework successfully addresses critical challenges in implementing secure and auditable distributed computing systems through its innovative three-layer architecture and hybrid consensus mechanism. The implementation demonstrates the feasibility of combining permissioned blockchain technology with sophisticated resource allocation schemes to support complex workflow orchestration while maintaining Byzantine fault tolerance.

Several promising directions emerge for future research and development of the TrustMesh framework. A key area of investigation lies in optimizing the framework's scalability through Sawtooth's dynamic consensus capabilities. The current implementation relies on PBFT consensus, which provides strong Byzantine fault tolerance but may impact performance in larger networks. Future work could explore automatically transitioning to Proof of Elapsed Time consensus as networks scale, effectively trading Byzantine fault tolerance for crash fault tolerance when appropriate for improved performance.

The framework could also benefit from enhancements to its network registry architecture, currently centralized within control nodes. Investigation into distributed storage technologies could provide pathways to decentralize the registry while maintaining the performance and security benefits of local image storage. Furthermore, integration with cloud resources through intelligent autoscaling mechanisms would enable TrustMesh to dynamically expand its computational capacity during high-demand periods while preserving security properties.

As IoT deployments continue to grow in scale and complexity, the need for secure, efficient, and trustless distributed computing frameworks becomes increasingly critical. TrustMesh lays the foundation for addressing these challenges through its novel architecture and security guarantees. The proposed future enhancements would not only expand its capabilities but also advance the broader field of blockchain-enabled computing, ultimately contributing to more resilient and scalable IoT ecosystems.

\section*{Acknowledgment}

This research was supported by The University of Melbourne’s Research Computing Services and the Nectar Research Cloud, a collaborative Australian research platform operated by the Australian Research Data Commons (ARDC) through NCRIS funding.

\bibliographystyle{IEEEtran}
\bibliography{IEEEabrv,references}

\end{document}